# Increasing the Number of Underrepresented Minorities in Astronomy
# Executive Summary
*An Astro2010 State of the Profession Position Paper*
March 2009

**Authored by**: The AAS Committee on the Status of Minorities in Astronomy (CSMA),
**with endorsement from:**
National Society of Hispanics in Physics - David J. Ernst, Pres. (Vanderbilt University),
Marcel Agueros (Columbia University), Scott F. Anderson (University of Washington), Andrew Baker (Rutgers University), Adam Burgasser, (Massachusetts Institute of Technology), Kelle Cruz (Caltech), Eric Gawiser (Rutgers University), Anita Krishnamurthi (University of Maryland, College Park), Hyun-chul Lee (Washington State University), Kenneth Mighell (NOAO), Charles McGruder (Western Kentucky University), Dara Norman (NOAO), Philip J. Sakimoto (University of Notre Dame), Kartik Sheth (Spitzer Science Center), Dave Soderblom (STScI), Michael Strauss (Princeton University), Donald Walter (South Carolina State University), Andrew West (MIT)
UW Pre-Map staff - Eric Agol (Faculty Project Leader), Jeremiah Murphy, Sarah Garner, Jill Bellovary, Sarah Schmidt, Nick Cowan, Stephanie Gogarten, Adrienne Stilp, Charlotte Christensen, Eric Hilton, Daryl Haggard, Sarah Loebman Phil Rosenfeld, Ferah Munshi (University of Washington)

Primary Contact
Dara Norman
NOAO
950 N. Cherry Ave
Tucson, AZ 85719
dnorman@noao.edu,
520-318-8361

Promoting racial and ethnic diversity is critically important to the future success and growth of the field of astronomy. The raw ability, drive and interest required to excel in the field is distributed without regard to race, gender, or socioeconomic background.  By not actively promoting diversity in our field we risk losing talented people to other professions (or losing them entirely), which means that there will be astronomical discoveries that simply won't get made.  There is demonstrated evidence that STEM fields benefit from diverse perspectives on problems that require more complex thought processes.[1] This is especially relevant to a field like astronomy where more and more work is being done collaboratively.

The lack of notable growth in African American, Hispanic, and Native American representation in astronomy indicates that the "pipeline" for these individuals is systemically leaky at critical junctures. Substantially more effort must be directed toward improving the educational and career development of minorities to insure that these potential colleagues are supported through the process. However, simply recognizing that the pipeline is faulty is woefully inadequate. There must be very specific, targeted solutions to help improve the situation.  With this in mind, we offer two position papers addressing specific areas of improvement that we identify as (a) essential for any foreseeable progress in the field, and (b) attainable in the 2010-2020 decade. These position papers focus primarily on African Americans, Hispanics, and Native Americans. Although we do not directly address issues of Asian Americans, Pacific Islanders, and other groups, many of the recommendations made here can be adapted to address issues faced by these groups as well. We summarize the discussion of these papers as follows:

- **Undergraduate, Graduate, and Postdoctoral Levels (Paper I)**

A clear means to improvement is in tending to the educational transitions in which potential minority PhDs are lost. This includes the transition for minority physics and astronomy undergraduates into astronomy and astrophysics PhD programs, or similar careers beneficial to the astronomy enterprise[3]. Partnerships with minority-serving institutions (MSIs) can provide an effective and immediately attainable solution, as they not only provide critical stepping stones to the PhD, but also because the strength of these undergraduates in physics, engineering, and computer science offer promising avenues for engagement in instrumentation development, support, and large scale computing/data-mining. Early and continuous research engagement is critical to this vision, in which the federally funded undergraduate research internship programs (e.g., NSF REU), and national centers and observatories (e.g., NOAO) play a vital role.

---

[1] Chubin, D.E. and Malcolm, S.M., "Making a Case for Diveristy in STEM Fields", Inside Higher Education, October, 6, 2008 and references within.  See www.insidehighered.com/views/2008/10/06/chubin.

[3] By "astronomy enterprise" we mean Astronomy and related fields that include Physics, engineering and computer science. See the section on "Enhance recruitment through physics and engineering".

- **K-12 Education and Public Outreach (Paper II)**

To keep our technological workforce strong in the next decade and beyond, we must proactively increase opportunities for minority students in STEM areas well before they reach the undergraduate level. The astronomical community can and should play a critical role in supporting these opportunities because of Astronomy's wide spread appeal and inspirational nature. Our community must improve its efforts to develop and sustain education opportunities for today's minority elementary, middle, and high schools students, in order to attract, recruit, and retain them in astronomy and related disciplines, and insure that there continues to be a well qualified pool of undergraduate and graduate students from which to recruit. Only by actively engaging these communities can the U.S hope to continue its leadership in astronomical discovery and knowledge.

These position papers were prepared with the intent that several study groups for the State of the Profession will take interest in them, with specific attention to the study groups on Demographics (DEM) and Education and Public Outreach (EPO). These submitted papers do not include their appendices, which can be found on the AAS CSMA's website at: http://csma.aas.org/events.html .

**1. <u>Statement of the Problem:</u>**
The vast under-representation of minorities in astronomy remains a staggering challenge despite at least two decades[KGS1] of awareness of the issue. Let us briefly examine the challenge quantitatively.

Over the past 20 years, the absolute number of PhDs awarded annually to underrepresented minorities in the field has grown slightly, from approximately 3±1 in 1988 to approximately 5±1 per year in recent years. The corresponding proportion of minority PhDs has been roughly flat at 2-4% of the total (see Paper I). During this same time period, the proportion of underrepresented minorities in the U.S. population grew by 33%, from 20.9% in 1988 to 27.0% in 2009 (data from US Census). Consequently the relative underrepresentation of these groups in astronomy and astrophysics has been steadily worsening.

What must be done to achieve parity in the production of underrepresented minorities in the field? As of 2004, the 50 PhD-granting institutions in astronomy and astrophysics counted 652 permanent (i.e. tenure-stream) faculty (Nelson & Lopez 2004). Multiply this number by a factor of ~4 to account for permanent jobs at non-PhD institutions and national centers, and we have that there are roughly 2600 permanent astronomy and astrophysics positions in the U.S. If an individual holding one of these positions occupies it for typically 30-35 years, then approximately 3% of the permanent jobs turn over every year, or about 75 positions. This is consistent with the recent estimate from the AAS Employment Committee, which found that on average 75±15 permanent jobs open up in the U.S. every year. Thus to achieve parity in the number of minorities entering the stream of permanent astronomy and astrophysics positions, the community must in the coming decade increase the number of minority PhDs from 5 per year to 20. And if the same attrition rates apply to these individuals as in the field overall (the AAS Employment Committee estimates that roughly 40% of astronomy and astrophysics PhDs end up in permanent jobs), then this number becomes 50. *In other words, the absolute number of* *minority PhDs produced annually must increase by a factor of at least 5-10 in the coming decade.* At this aggressive pace, the field can achieve parity overall in 30-35 years.

It is moreover essential to realize the scale of PhD production problem. Currently there are fewer than 20 minority faculty at astronomy and astrophysics PhD-granting institutions (Nelson & Lopez 2004). If only these individuals are actively involved in increasing minority PhD numbers, each would need to produce 2-3 minority PhDs per year to achieve these target goals. Vastly increasing the number of minority PhDs in the coming decade will have to be the purview of the entire astronomy and astrophysics community.

At the same time, the issue of under-representation in any STEM field, including astronomy, is not merely about achieving target demographics. At its heart, it concerns the tremendous social issues related to the perpetual educational and economic inequity in our nation. Its solution should address the flat (if not declining) scientific and technical ability of the general workforce (see Paper II), upon which our nation's economic strength and global leadership will depend on in the next decade and beyond. Astronomy will not be immune to these deficiencies, and is especially dependent in areas of instrumentation development and the computational challenges associated with mining and serving vast datasets. Improving minority participation will help meet these challenges by providing a sufficient domestic, technically-trained workforce to complete, support, and lead the advancement of astronomy.

2. **The primary recommendations of these papers are:**

*For Colleges, Universities, and National Centers/Observatories:*
- Commit to engaging local under-served minority communities, to encourage interest in and appreciation of math and science, and retain their interest in astronomy.
- Develop "horizontal" and "vertical" partnerships with MSIs. Form equal-stakes partnerships in research, funding, and development that reflect the mutual synergies of intellectual contribution.
- Develop internship programs that connect minority students to mentored research engagement with scientific and/or engineering staff.

*For Funding Agencies:*
- Substantially expand funding for programs that specifically forge linkages between MSIs and research universities.
- Provide funding incentives for broadening participation of underrepresented minorities in mission critical ways by including this in the funding criteria (e.g., NSF's "broader impacts" criterion).
- Provide opportunities for continuity of funding for astronomy education programs by establishing Federal funding cycles in the same way that research is funded.

*For Professional Societies:*
- Play a lead role in aggressively identifying exceptionally effective K-12 outreach programs and work to see them adopted widely, particularly in under-served communities.

- Create a professional network to better link potential candidates (at all levels) with potential employers and programs.

### *3. Other Issues of Importance*

The targeted solutions presented in these position papers are in no way meant to be the "silver bullets" that fix the numerous challenges related to diversity and racial equity, even within our field. Indeed, there are several other areas of critical importance that must be addressed in the coming decade, which for the sake of brevity and focus, are not presented in our papers. Efforts must be made on several fronts to ensure a more comprehensive growth in minority representation. These include, but are not limited to,

- Advocating a greater investment in middle and high school science education, specifically those in dense minority centers. This includes encouraging strong undergraduates to pursue teaching careers, developing stronger science preparation for certification, and strongly supporting better teacher salaries.
- Pre- and post-PhD mentorship for career development, to insure that the careers of minority investigators remain "on track" for faculty and research positions, and to provide "off-ramp" career options at important junctures (after graduation, after postdocs, etc.).
- A greater awareness of cultural issues that amount to professional obstacles for minorities in the traditional pipeline. These include the very realistic importance for many to stay in their home regions, for schooling and ultimately for their careers.